\documentclass[namedreferences]{solarphysics}
\usepackage[hyperref,optionalrh,solaromanenum]{spr-sola-addons} 
\usepackage{graphicx}                    
\usepackage{amssymb}                    
\usepackage{color,xcolor}                      
\usepackage{colortbl} 
\usepackage{multirow}
\usepackage{soul}
\usepackage{ulem}
\soulregister\cite7
\soulregister\citealp7
\soulregister\ref7

\chardef\us=`\_

\begin{document}

\begin{article}
\begin{opening}

\title{Origin of the Chromospheric Umbral Waves in Sunspots}

\author[addressref={ad1,ad2},email={zhangxinsheng@ynao.ac.cn}]{\fnm{Xinsheng }\lnm{Zhang}}
\author[addressref={ad1,ad3},corref,email={yanxl@ynao.ac.cn}]{\fnm{Xiaoli }\lnm{Yan}\orcid{0000-0003-2891-6267}}
\author[addressref={ad1,ad3},email={zkxue@ynao.ac.cn}]{\fnm{Zhike }\lnm{Xue}\orcid{0000-0002-6526-5363}}
\author[addressref={ad1,ad3}]{\fnm{Jincheng }\lnm{Wang}}
\author[addressref={ad1,ad3}]{\fnm{Zhe }\lnm{Xu}}
\author[addressref={ad4,ad5}]{\fnm{Qiaoling }\lnm{Li}}
\author[addressref={ad1,ad2}]{\fnm{Yang }\lnm{Peng}}
\author[addressref={ad1,ad2}]{\fnm{Liping }\lnm{Yang}}

%
\runningauthor{X. Zhang et al.}
\runningtitle{The Origin of Umbral Waves}

\address[id={ad1}]{Yunnan Observatories, Chinese Academy of Sciences, Kunming 650216, China}
\address[id={ad2}]{University of Chinese Academy of Sciences, Beijing 100049, China}
\address[id={ad3}]{Yunnan Key Laboratory of Solar Physics and Space Science, Kunming 650216, China}
\address[id={ad4}]{Department of Physics, Yunnan University, Kunming 650091, China}
\address[id={ad5}]{Department of Astronomy, Key Laboratory of Astroparticle Physics of Yunnan Province, Yunnan University, Kunming 650091, China}
\begin{abstract}
Oscillations are ubiquitous in sunspots and the associated higher atmospheres. However, it is still unclear whether these oscillations are driven by the external acoustic waves (p-modes) or generated by the internal magnetoconvection. To obtain clues about the driving source of umbral waves in sunspots, we analyzed the spiral wave patterns (SWPs) in two sunspots registered by IRIS  Mg\uppercase\expandafter{\romannumeral2} 2796 Å slit-jaw images. By tracking the motion of the SWPs, we find for the first time that two one-armed SWPs coexist in the umbra, and they can rotate either in the same or opposite directions. Furthermore, by analyzing the spatial distribution of the oscillation centers of the one-armed SWPs within the umbra (the oscillation center is defined as the location where the SWP first appears), we find that the chromospheric umbral waves repeatedly originate from the regions with high oscillation power and most of the umbral waves occur in the dark nuclei and strong magnetic field regions of the umbra. Our study results indicate that the chromospheric umbral waves are likely excited by the p-mode oscillations.

\end{abstract}
%

\end{opening}

%
\section{Introduction}
	\label{sec:intro}

Various oscillation phenomena exist in the atmosphere of sunspots. In the photosphere, the oscillations of the sunspot umbra usually have a dominant period of 5 minutes \citep{Bhatnagar+etal+1972, Soltau+etal+1976} and a weak signal of 3-minute oscillations \citep{Kobanov+etal+2011, Sych+etal+2020}. As the atmospheric altitude increases, the amplitude of the 5-minute oscillations gradually decreases, while the amplitude of the 3-minute oscillations gradually increases \citep{Chae+etal+2017, Felipe+etal+2018}. Up to the chromospheric height, the 3-minute oscillations dominate in the umbra \citep{Beckers+etal+1972}. Compared with the 3-minute oscillations, 5-minute oscillations become extremely weak \citep{Priya+etal+2018, Sych+etal+2020}. Umbral waves with a period of about 3 minutes persist in the transition region and the corona \citep{DeMoortel+etal+2002, Sych+etal+2009, Tian+etal+2014}, and the energy that they carry is eventually dissipated in the corona \citep{DeMoortel+etal+2003, Jess+etal+2012}. Additionally, a radially outward propagating wave pattern, called running penumbral waves, was observed in the penumbra of sunspot \citep{Giovanelli+1972, Zirin+etal+1972, Musman+etal+1976}. Whether in the photosphere, chromosphere, or higher solar atmosphere, the dominant oscillation period of the running penumbral waves is generally about 5 minutes \citep{Brisken+etal+1997, Priya+etal+2018}. According to the current understanding, the umbral waves and running penumbral waves are different manifestations of slow magnetoacoustic waves that propagate from the photosphere to the corona \citep{Khomenko+etal+2015, Li+etal+2020}.

In addition to the directly observed wave pattern, fast-moving photospheric waves have been detected using helioseismic techniques \citep{Zhao+etal+2015}, and numerical simulations have shown that such waves are generated by sources located at different depths between approximately 1 Mm and 5 Mm below the sunspot \citep{Felipe+etal+2017}. Recently, \cite{Yuan+etal+2023} detected transverse waves in the chromospheric umbral fibrils. Studying these different waves can help us understand the thermal and magnetic structure of sunspots \citep{Shibasaki+2001, Yuan+etal+2014}.

Despite the increasing literature, the origin of umbral waves in sunspots remains controversial. Currently, the external p-mode oscillations and the internal magnetoconvection are believed to be two possible driving sources of the umbral waves \citep{Khomenko+etal+2015}. 

\cite{Braun+etal+1987} and \cite{Penn+etal+1993} found that sunspots absorb some of the incident energy of p-modes. The wave energy absorption coefficient is determined by calculating the ratio of the outgoing and incoming waves. The p-modes absorption coefficient increases with the increase of the horizontal wavenumber, up to a maximum of about 50\% \citep{Braun+etal+1987, Bogdan+etal+1993}. Observations show that the p-modes absorption coefficient is roughly proportional to the magnetic flux density \citep{Braun+etal+1988, Braun+etal+1990}. The dark nuclei regions in the umbra, where the magnetic field is strongest \citep{Thomas+etal+2004}, have the highest absorption coefficient for p-modes. Some observational evidence supports the p-modes driving, including the p-modes absorption coefficient \citep{Penn+etal+1993, Braun+1995}, the similarity of the frequency spectrum and amplitude modulation with the quiet Sun \citep{Zhao+etal+2013, KrishnaPrasad+etal+2015}. The interaction of the p-modes with strong magnetic fields in a sunspot could excite the magnetoacoustic modes \citep{Cally+etal+1997, Cally+etal+2003, Khomenko+etal+2013}. The magnetic field in sunspots acts as a filter, preferentially allowing the p-modes from a narrow range of incident directions to pass through \citep{Schunker+etal+2006}.  It is important to note that this p-modes absorption is not real absorption, but is partially converted into slow magnetoacoustic waves and Alfv\'{e}n waves, which disappear into the solar interior under the guidance of the sunspot magnetic field \citep{Cally+etal+2003, Crouch+etal+2005, Cally+etal+2016}. Therefore, if the umbral waves are driven by the p-modes, then a large number of umbral waves should be observed above the dark nuclei regions with high absorption coefficients.

In addition, it has been proposed that waves can be excited by magnetoconvection occurring inside a sunspot \citep{Moore+1973, Lee+1993, Jacoutot+etal+2008}. Observational evidence suggests that the 3-minute oscillation power is enhanced above the light bridge and umbral dots of sunspots \citep{Jess+etal+2012, Yurchyshyn+etal+2015, Chae+etal+2017}. \cite{Cho+etal+2019} reported four umbral oscillation events that correspond to umbral dots directly below them, which further supported the internal excitation mechanism. Thus, a contradiction arises: do the umbral waves concentrate in the dark nuclei regions of the umbra, or in the brighter regions such as the light bridge and the umbral dots? In this way, the problem of finding the driving source of the umbral waves is transformed into a problem of exploring the spatial relationship between the umbral waves and the fine structures of the umbra.

With the deployment of more advanced astronomical instruments, the fine structure of sunspot umbral waves has been detected. As the umbral waves propagate upward along the magnetic field lines, the wave pattern observed at some heights exhibits a clear clockwise (counter-clockwise) rotation. \cite{Sych+etal+2014} discovered that the umbral waves exhibited a two-armed spiral wave pattern (SWP) in the ultraviolet band. Subsequently, \cite{Sych+etal+2021} performed a dynamical study of the 3-minute wavefronts. \cite{Su+etal+2016} found one-armed and multi-armed SWPs with rotational directions that can be changed within the chromospheric umbra. \cite{Felipe+etal+2019} detected a two-armed SWP in the Doppler velocity maps of the high photosphere. \cite{Kang+etal+2024} identified 241 SWPs from 140 sunspots, most of which have one spiral arm structure. They found that the properties of the SWPs were independent of the hemisphere, latitude, and sunspot size.

To explain this phenomenon of spiral arm structures, \cite{Kang+etal+2019} proposed a theoretical model in which they suggested that SWPs are produced by the superposition of non-zero azimuthal modes driven at 1600 km below the photosphere. The one-armed SWP is generated by the superposition of the slow-body sausage ($m = 0$) and kink ($m = +1$ or -1) modes, while the two-armed SWP is produced by the superposition of the slow-body sausage ($m = 0$) and fluting ($m = +2$ or -2) modes. The number and rotation direction of the spiral arms are determined by the absolute value and sign of the non-zero-m, respectively. Therefore, the observed apparent SWPs are not caused by the wave propagation in the azimuthal direction, but by the fact that the waves with the same phase arrive later at increasing distance from the axis of the waveguide magnetic flux tube, which results in a delayed rotation of the wave patterns and forms the trailing spiral arm structures \citep{Madsen+etal+2015, Cho+etal+2015, Kang+etal+2019, Cho+etal+2020}. Subsequently, observational studies have shown that sunspots indeed present multiple concurrent magnetohydrodynamic (MHD) modes \citep{Jess+etal+2017, Albidah+etal+2021, Albidah+etal+2022, Albidah+etal+2023}. Recently, \cite{Wu+etal+2021} proposed a complete dispersion relation that includes magnetic twist and the kink mode of $ m=-1 $, and further showed that magnetic twist has little influence on the morphology of the SWPs in the frame of linear perturbation analysis.

\cite{KrishnaPrasad+etal+2017} computed the amplitudes of 3-minute umbral oscillations at different heights of the solar atmosphere and found that the oscillations at the chromosphere (2796 Å) have large amplitudes. We found that the IRIS 2796 Å images clearly show the SWPs of the umbral waves without requiring any phase-velocity or band-pass filtering. However, no one has yet studied the fine structure of the umbral waves based on 2796 Å imaging data. Thus, in this paper, we analyze the spatial distribution of the oscillation centers of the SWPs within the umbrae to reveal the spatial relationship between the oscillation centers and the dark nuclei/umbral dots based on the imaging data at 2796 Å. The oscillation center is defined as the location where the SWP first appears, which is also the location of the origin of the umbral wave. We present the observational aspects of the data in Section \ref{sec:obs}, then present our analysis and results in Section \ref{sec:result}, and finally discuss and summarize the main results of the research in Section \ref{sec:conclusion}.

\begin{figure}
	\centering    
	\includegraphics[width=1\linewidth]{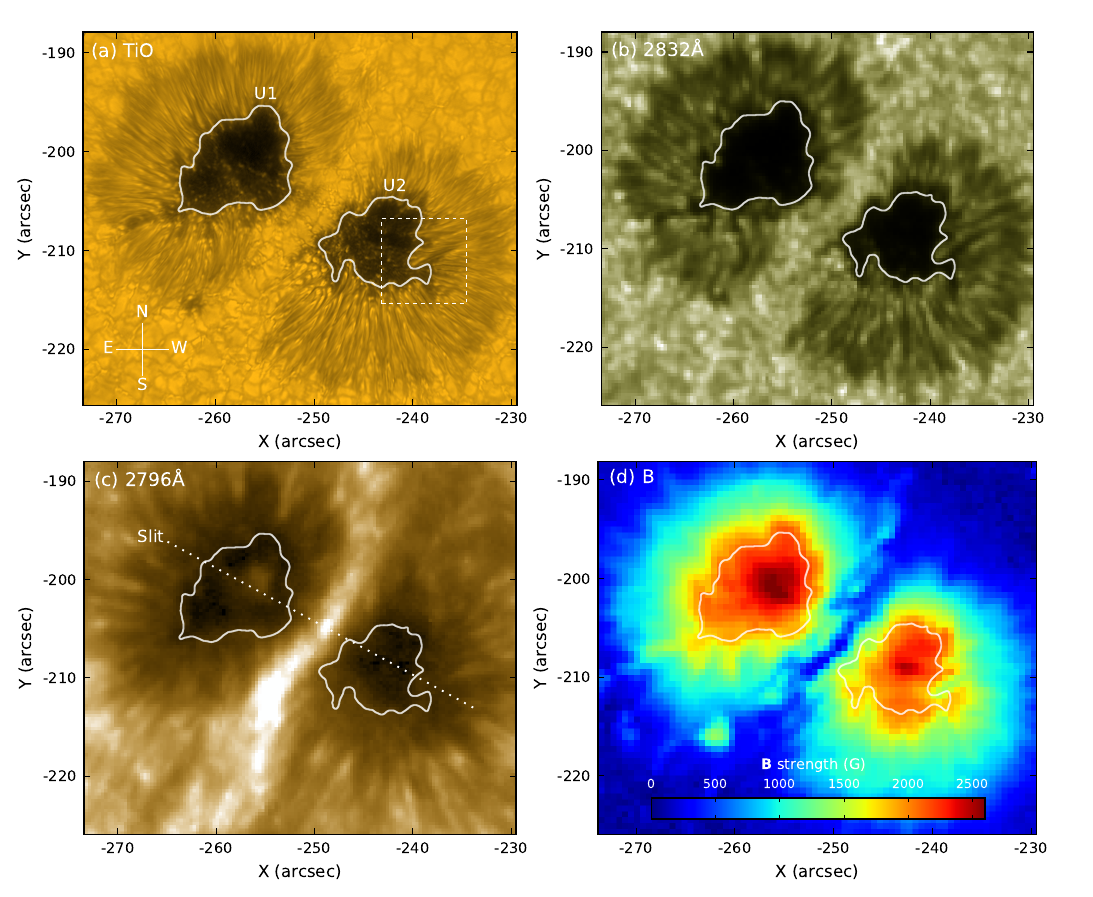}
	\caption{The appearance of two sunspots in the active region NOAA 13023 on 29 May 2022 at 17:51:40 UT. Panel (a): TiO image, in which the umbra regions of two sunspots are labeled as U1 and U2, respectively. The \textit{white contours} outline the umbral boundaries. The square region marked by the \textit{dashed line} is shown again in the inset of Figure \ref{Figure_2}(b). Panel (b): 2832 Å image. Panel (c): 2796 Å image, with a virtual slit marked by the \textit{dotted lines} for further analysis, as shown in Figure \ref{Figure_2}. Panel (d): Total magnetic field strength image. }	
	\label{Figure_1}	
\end{figure}

\section{Observations}
\label{sec:obs}

On 29 May 2022, two homopolar sunspots in active region NOAA 13023 (-206$^\prime$$^\prime$, -250$^\prime$$^\prime$) were observed by the Interface Region Imaging Spectrograph \citep[IRIS:][]{DePontieu+etal+2014}. In this observation, IRIS scanned the sunspots from 17:00 UT to 22:45 UT in an “eight-step raster” mode, with a step size of 2$^\prime$$^\prime$. Since the spectrographic slit was pointed entirely outside the sunspots, we only consider images taken by the slit-jaw imager (SJI) in two channels, 2796 Å and 2832 Å, with pixel scales of 0.3327$^\prime$$^\prime$ and time cadences of 20 s and 122 s, respectively. The 2796 \r{A} and 2832 \r{A} channels mainly capture the plasma present in the chromosphere and the photosphere, respectively \citep{DePontieu+etal+2014}. The data for each band were processed to the level 2 standard and self-aligned. Each IRIS SJI Flexible Image Transport System (FITS) file header provides the arcsecond coordinates of the field-of-view center relative to the solar disk center, which allows us to accurately align the IRIS SJI 2796 Å or 2832 Å images with the HMI continuum images. We have removed the data seriously affected by energetic particles and divided the original $\sim$ 6 hours of continuous observations into four-time intervals (17:00-17:55 UT, 18:15-19:30 UT, 20:10-21:06 UT, and 21:20-22:45 UT). In addition, We also used the SJI 2796 Å imaging data from IRIS obtained on 30 May 2022 between 11:10 and 11:15 UT for the same active region to show the SWPs of the umbral waves.

We used 720 s vector magnetograms acquired by the Solar Dynamics Observatory/Helioseismic and Magnetic Imager \citep[SDO/HMI:][]{Schou+etal+2012} at four time intervals on May 29, and we converted the data into the latitudinal  $(B_{p})$, longitudinal $(B_{t})$, and radial $(B_{r})$ components. Moreover, the Goode Solar Telescope (GST) located at the Big Bear Solar Observatory (BBSO) provided the joint observational data. We used the speckle-reconstructed TiO (7057 Å) broadband filter images \citep{Cao+etal+2010} to extract the sunspot umbral boundaries. The data have a time cadence of 30 s and a pixel scale of 0.034$^\prime$$^\prime$. Because of the unstable seeing conditions during the observation, we selected the best-quality TiO image in each time interval and used the Scale-Invariant Feature Transform algorithm \citep[SIFT:][]{Lowe+2004, Ji+etal+2019} to align the TiO images and the HMI continuum images with sub-pixel accuracy.

\begin{figure}
	\centering    
	\includegraphics[width=1\linewidth]{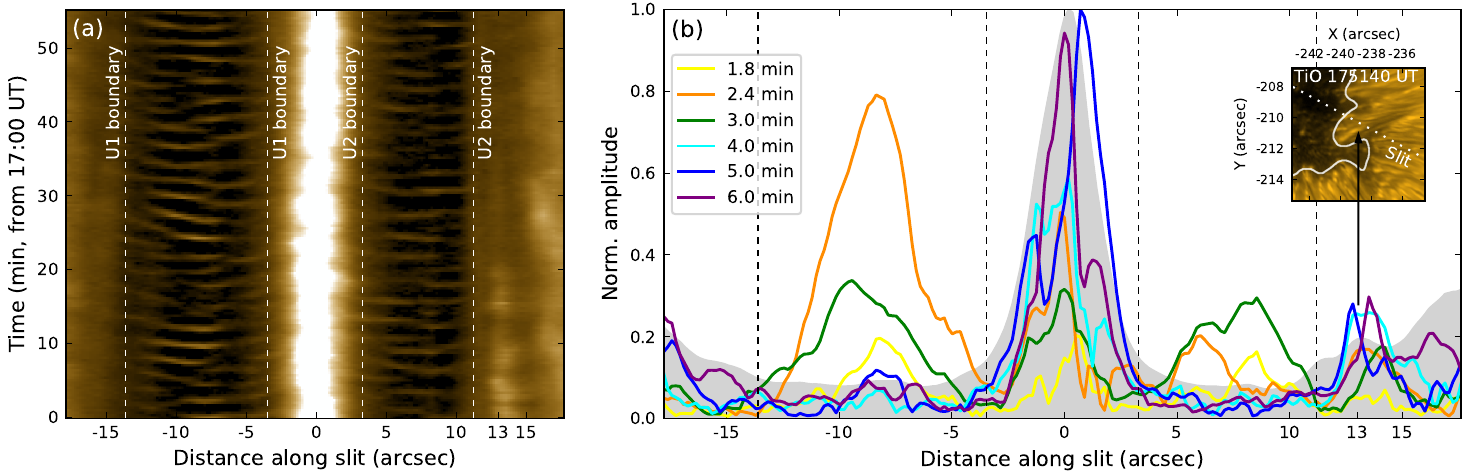}
	\caption{Panel (a): The time-distance diagram of the 2796 \r{A} intensity obtained from the slit in Figure \ref{Figure_1}(c). Panel (b): The distribution of narrowband oscillation amplitudes along the SJI 2796 \r{A} slit is represented by various colors, each corresponding to different oscillation periods. The \textit{gray shading} illustrates the variation of the mean intensity. The inset corresponds to the field of view shown by the \textit{dashed box} in Figure \ref{Figure_1}(a). The \textit{dashed lines} in panels (a) and (b) mark the boundaries of the umbra.}	
	\label{Figure_2}	
\end{figure}
\section{Analysis and Results} 
\label{sec:result}
\subsection{Oscillations of the Umbra, Penumbra, and Granulation}

Figure \ref{Figure_1} shows the appearance of the active region NOAA 13023 obtained on 29 May 2022. Two sunspots with the same magnetic polarity are closely spaced and separated by a narrow granulation channel, and the umbrae are exposed on the side facing the granulation channel. The umbrae of two sunspots are labeled as U1 and U2, respectively. It can be clearly seen that the shapes of the two umbrae are markedly different (see the white contours). The mean magnetic field strengths of U1 and U2 were 2130 G and 1924 G, respectively, during the four time intervals on 29 May. As time evolved, both sunspots were in a decay phase and rotated counterclockwise. \cite{Peng+etal+2024} conducted a detailed study of the decay phase of the active region NOAA 13023.

We first determined the frequency distribution of oscillatory signals in different regions. Figure \ref{Figure_2}(a) shows the time-distance diagram of the 2796 Å intensity along the slit marked in Figure \ref{Figure_1}(c). The bright region in the center of the image is intentionally overexposed to highlight the details of the umbral waves. There are many slightly inclined striations in the umbra, which are called the umbral waves. The umbral wave first appeared inside the umbra, then gradually moved toward both ends of the slit, and finally became invisible at the umbral boundaries. It should be emphasized that the horizontal motion of the umbral wave is not due to the wave propagation in the azimuthal direction, but results from the apparent phenomenon that the wave with the same phase arrives later as the distance from the waveguide magnetic flux tube axis increases \citep{Madsen+etal+2015, Cho+etal+2015, Kang+etal+2019, Cho+etal+2020}. For each slit position in Figure \ref{Figure_2}(a), a Hanning window is applied to the time series, followed by the computation of the amplitude spectrum via a Fast Fourier Transform. Therefore, Figure \ref{Figure_2}(b) illustrates the variation of the oscillation amplitudes across different frequency bands along the slit, normalized relative to the peak value. By comparing the frequency distributions of the oscillation signals in different regions, we found that the umbrae are dominated by oscillations with periods of 1.8 to 3.0 minutes, while the penumbras and granulation are dominated by oscillations with periods of 4.0 to 6.0 minutes. The dominant oscillation frequency and mean oscillation amplitude of U1 are higher than those of U2, and the oscillation signals are relatively weak at the umbral boundaries. We obtained the same observational characteristics in the slits determined in other directions. It should be acknowledged that Figure \ref{Figure_2}(b) only roughly reflects the differences in the frequency distribution of the oscillation signals in different regions. Accurate calculation of the dominant oscillation periods in different regions is beyond the scope of our study. Additionally, it is interesting to note that the oscillations are more intense at the slit position of 13$^\prime$$^\prime$, where a cluster of bright penumbral filaments is gradually crossing below (see the location indicated by the black arrow). In Figure \ref{Figure_2}(a), the brightness at the slit position of 13$^\prime$$^\prime$ decreases gradually with time, which is likely a chromospheric response to the penumbral bright filaments crossing the slit location.

\begin{figure}
	\centering    
	\includegraphics[width=1\linewidth]{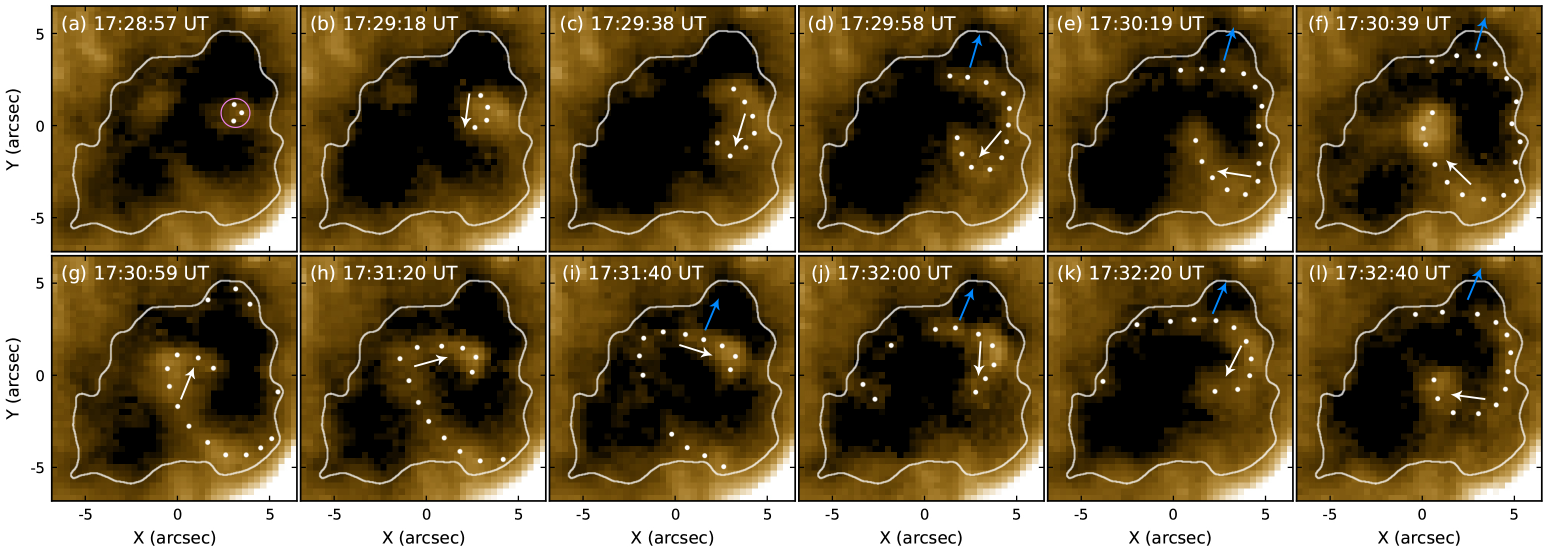}
	\caption{Formation of a clockwise rotating one-armed SWP within U1 in the SJI 2796 Å image. The \textit{white dotted curves} emphasize the propagation trajectory of the wavefront, and the \textit{white solid line} outlines the umbral boundary. The \textit{white} and \textit{blue arrows} show the wave-propagation directions. The \textit{violet circles} indicate the oscillation centers which is defined as the location where the SWP first appears. (An animation of this figure is available).}	
	\label{Figure_3}	
\end{figure}

\subsection{Spiral Wave Patterns (SWPs) in the Umbra}

There are abundant and clear SWPs within the umbrae of two sunspots. In Figure \ref{Figure_3}, we investigate the evolution of a one-armed SWP in U1. A bright circular wavefront emerges on the western side of the umbra (see the violet circle) and then evolves into a one-armed SWP in a clockwise direction (see the white arrows). During the subsequent clockwise rotation, the head of the wavefront remains near the oscillation center, while the tail of the wavefront expands radially outward (see the blue arrows) and shows signs of fragmentation (e.g., panel (g) and panel (i)). Within nearly 4 minutes, this one-armed SWP rotated by about 450° around the oscillation center. Before this, \cite{Kang+etal+2019} proposed that the one-armed SWP was produced by the superposition of the slow-body sausage ($m = 0$) and the kink modes ($m = +1$ or -1). However, by comparing with the evolution process of the kink mode in the theoretical models \citep{Kang+etal+2019, Wu+etal+2021}, we find that the one-armed SWP in Figure \ref{Figure_3} seems to be solely composed of an azimuthally non-symmetric mode ($m = -1$). It should be emphasized that the kink mode in the sunspot considered here is a longitudinal wave associated with slow magnetoacoustic waves \citep{LópezAriste+etal+2016, Jess+etal+2017}.

\begin{figure}
	\centering    
	\includegraphics[width=1\linewidth]{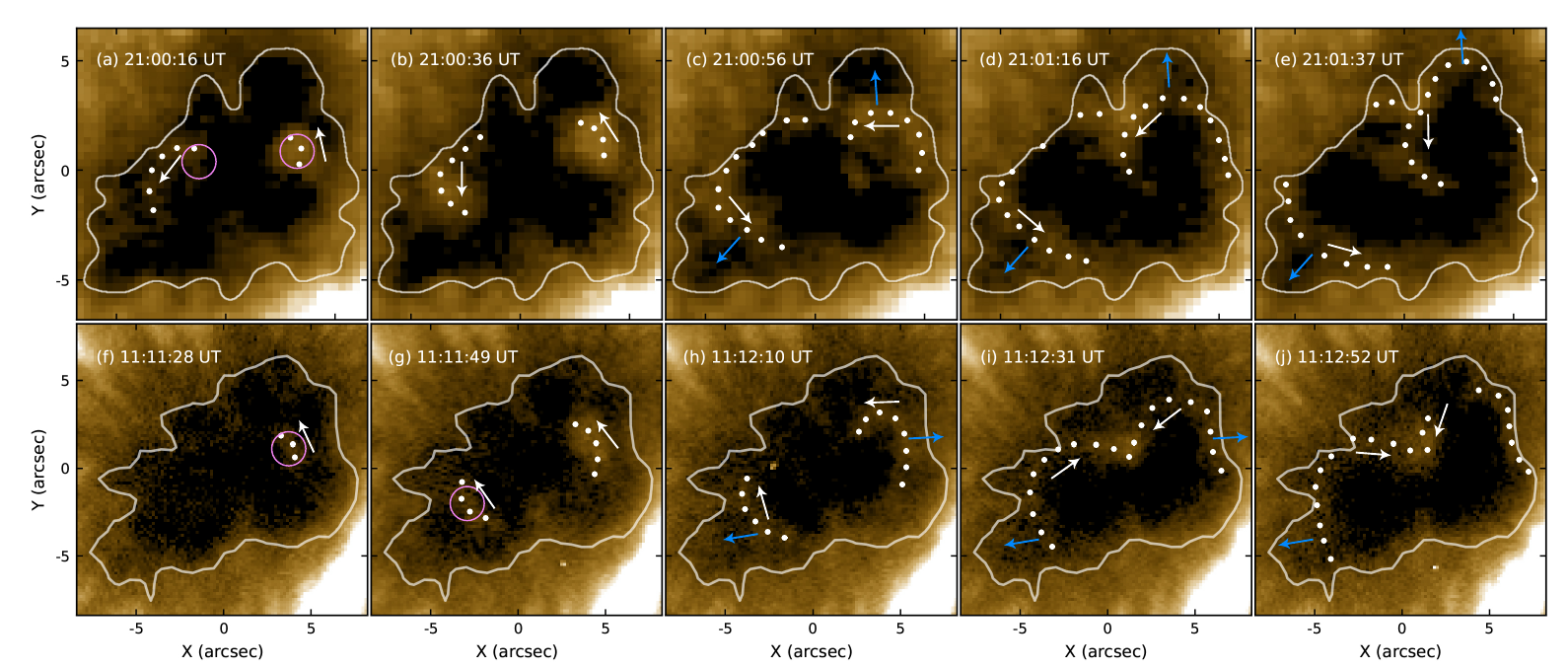}
	\caption{Similar to Figure \ref{Figure_3}, but the \textit{top panel} (from left to right) shows two one-armed SWPs rotating in the same direction within U1. The \textit{bottom panel} (from left to right) shows two one-armed SWPs rotating in opposite directions within U1. (An animation of this figure is available).}	
	\label{Figure_4}	
\end{figure}

\begin{figure}
	\centering    
	\includegraphics[width=1\linewidth]{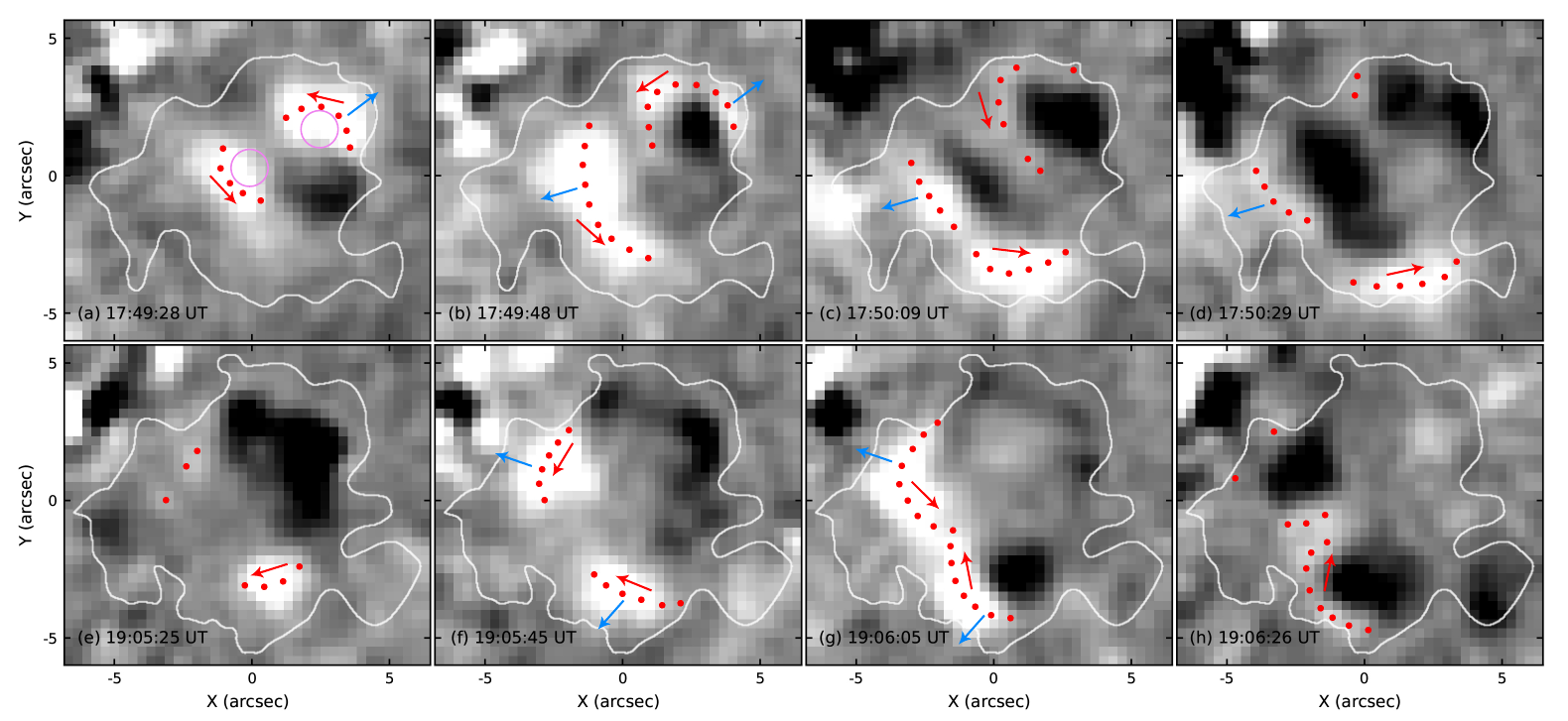}
	\caption{Panels (a-d): Two one-armed SWPs rotating in the same direction within U2 (seen in the running difference image at SJI 2796 Å). The \textit{red dotted curves} emphasize the propagation trajectory of the wavefronts, and the \textit{white solid line} outlines the boundary of the umbra. \textit{Red} and \textit{blue arrows} show the wave-propagation direction. The \textit{violet circles} indicate the oscillation centers. Panels (e-h): Two counterrotating one-armed SWPs within U2 (whose spatial position of the oscillation centers is undetermined and thus not marked). (An animation of this figure is available).}	
	\label{Figure_5}	
\end{figure}

More interestingly, we find for the first time that two one-armed SWPs coexist at different spatial locations within U1 or U2, as shown in Figures \ref{Figure_4} and \ref{Figure_5}. The top row of Figure \ref{Figure_4} (from left to right) shows two co-rotating one-armed SWPs within U1 (see the white arrows). In contrast, the bottom row (from left to right) shows two counter-rotating one-armed SWPs within U1, which meet at 11:12:31 UT (panel (i)) and then gradually dissipate. Similarly, in U2 (Figure \ref{Figure_5}), we find two one-armed SWPs that can rotate either in the same or opposite directions. To highlight the spiral arm structure, each panel in Figure \ref{Figure_5} shows a running difference image obtained using SJI 2796 Å data with a time interval of two frames, and has been smoothed over a width of three pixels. It is noteworthy that the simultaneous presence of two one-armed SWPs in the umbra is a rare phenomenon. We detected only five occurrences of this phenomenon in about five hours. The spatial positions and rotation directions of these two spiral arms are somewhat independent.

\begin{figure}
	\centering    
	\includegraphics[width=1\linewidth]{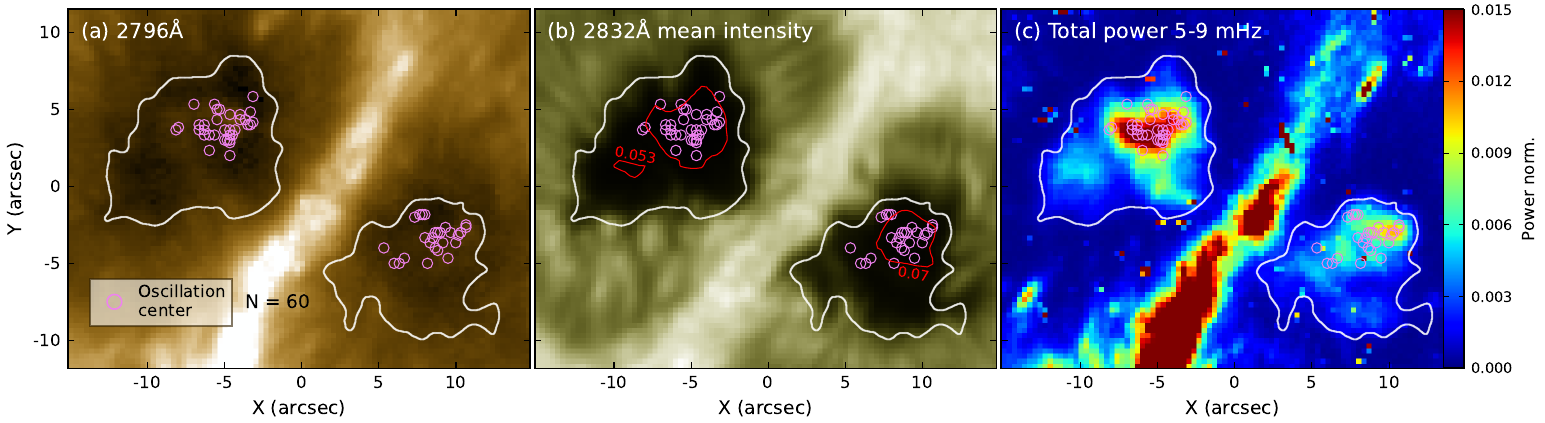}
	\caption{The spatial distribution of the oscillation centers of one-armed SWPs. The background image in panel (a) is from the IRIS SJI 2796 Å data. The \textit{violet circles} in panels (a)-(c) indicate the identified oscillation centers, and the \textit{white contours} outline the umbral boundaries. The background image in panel (b) is the mean intensity map obtained from SJI 2832 Å data at four time intervals on 29 May. \textit{Red contours} are drawn at levels of 0.053 and 0.07 within U1 and U2, respectively, to indicate the dark nuclear regions in the umbra (the contour thresholds are determined by visual inspection using the TiO band as a reference). Panel (c) shows the total oscillation power in the 5-9 mHz range computed from the SJI 2796 Å data at four time intervals on 29 May.}	
	\label{Figure_6}	
\end{figure}

\subsection{Spatial Distribution of Oscillation Centers}

To investigate the driving source of umbral waves, we identified 60 oscillation centers of one-armed SWPs within the umbrae of two sunspots. We excluded the multi-armed SWPs because their identification is somewhat subjective. In Appendix \ref{material}, we provide a detailed account of the start and end times of each one-armed SWP in a table. The mean lifetime of a one-armed SWP was $125 \pm 38$ s. The spatial distribution of the oscillation centers of one-armed SWPs is shown in Figure \ref{Figure_6}(a), where we note that the oscillation centers are not randomly distributed in the umbra, but rather are concentrated in specific regions within it. Figure \ref{Figure_6}(c) shows the total oscillation power map in the 5-9 mHz range, where red/blue colors indicate high/low oscillation power. We find that the oscillation centers are concentrated in regions with high oscillation power inside the umbra, implying that umbral waves are more likely to be triggered in regions of high oscillation power. 

Next, we look for the spatial relationship between the oscillation centers and the dark nuclei/umbral dots. Owing to the unstable seeing during the observation, we were unable to obtain high-quality TiO data continuously, so we used the IRIS SJI 2832 \r{A} images. Although the SJI 2832 \r{A} image cannot distinguish the umbral dot, it can identify the dark nuclei within the umbra. We averaged the 2832 \r{A} images over time to represent them for the four time intervals on 29 May. Figure \ref{Figure_6}(b) shows the mean intensity map at 2832 \r{A}, where the red contours approximately delineate the dark nuclei regions of the umbra. We used the TiO band as a reference to determine the threshold of the red contour lines based on visual inspection. The dark nuclei obtained by this method are inevitably different from the true dark nuclei, because the dark nuclei evolve gradually with time and the two sunspots rotate counterclockwise. However, it is undeniable that the brightness of the red contour regions is lower than the mean brightness of the umbra background. Studies have shown that dark nuclei have a strong magnetic field, where convection is almost completely suppressed \citep{Weiss+2002, Thomas+etal+2004}. In Figure \ref{Figure_6}(b), we find that the oscillation centers of the superposition are concentrated in the dark nuclei regions of the umbra. This indicates that most of the chromospheric umbral waves originate from the dark nuclei regions of the photosphere, rather than the brighter regions in the umbra. Therefore, our observations suggest that the chromospheric umbral waves are likely excited by the p-modes outside the sunspots.

\begin{figure}
	\centering    
	\includegraphics[width=1\linewidth]{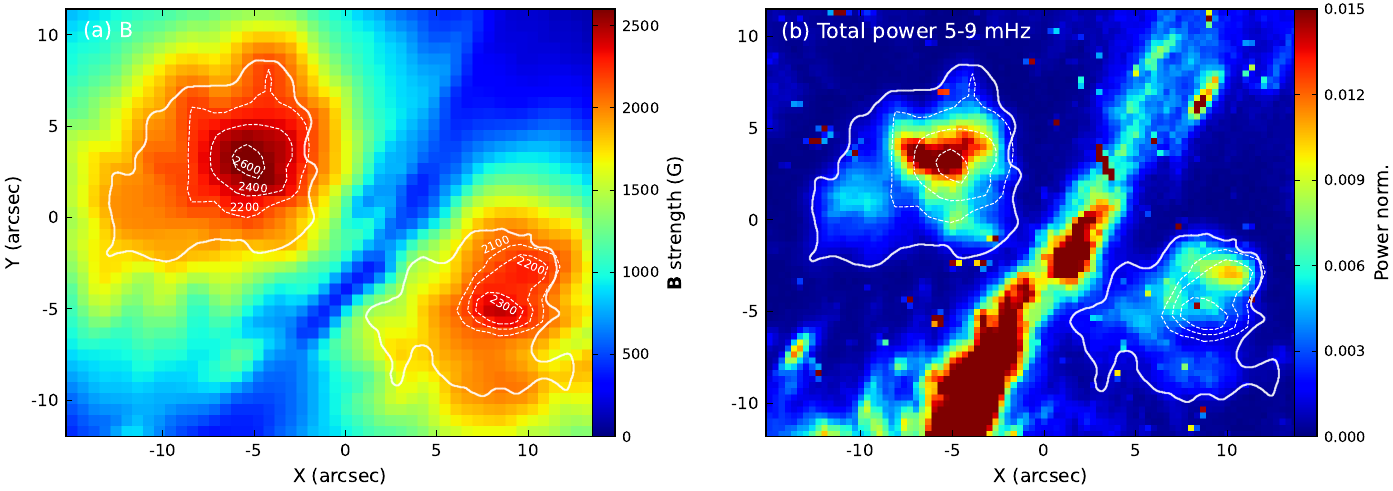}
	\caption{Panel (a): The mean magnetic field strength (0-2600 G) obtained from the 720 s vector magnetic field data at four time intervals on 29 May. Panel (b): The power maps of the 5-9 mHz total oscillations obtained from the SJI 2796 Å data for four time intervals on May 29, overlaid with the contours of the mean magnetic field strength. The \textit{white contours} in U1 are drawn at the levels of 2200, 2400, and 2600 G, while those in U2 are drawn at the levels of 2100, 2200, and 2300 G. The \textit{white solid lines} outline the umbral boundaries.}	
	\label{Figure_7}	
\end{figure}

Based on the transitivity of the spatial relations, the majority of regions with high oscillation power in the 5-9 mHz range should correspond to a strong magnetic field. In Figure \ref{Figure_7}, we show the maps of the mean magnetic field strength (the left panel) and the total oscillation power in the range of 5-9 mHz (the right panel). The superposed contours of the mean magnetic field strength in Figure \ref{Figure_7}(b) (see the white dashed lines) reveal that there is indeed a spatial relation between the locations of enhanced oscillations and the regions of strong magnetic field. This spatial correspondence relation is particularly evident in U1. In U2, although the region of the enhanced oscillations and the strong magnetic field region are both on the northwest side of the umbra, there is a spatial offset between the place where the chromospheric 5-9 mHz oscillation power is maximum and the place where the photospheric magnetic field is strongest. \cite{Wu+etal+2023} found that the energy of the 3-minute oscillations is more easily extended outward in the horizontal direction when slow magnetoacoustic waves propagate along the more inclined magnetic-field lines. Therefore, we believe that this spatial offset is due to the strong magnetic-field lines within the right sunspot extending toward the northwest as the atmospheric height increases.

\section{Discussion and Conclusions} 
\label{sec:conclusion}

In this study, we mainly use the IRIS SJI 2796 Å images to investigate the SWPs of umbral waves and explore the driving source of umbral waves by analyzing the spatial distribution of the origin of umbral waves.

Firstly, we observed more intense chromospheric oscillations above a cluster of bright filaments in the penumbral region of a sunspot, which reflects the inhomogeneity of the penumbral oscillations. We speculate that this is due to the more vertical magnetic field in the bright filaments (compared to the dark filaments), which enables running penumbral waves propagating along the bright filaments to reach the chromosphere more easily. \cite{Sych+etal+2020} found that regions with enhanced 5-minute oscillation power in the chromospheric penumbra have a filamentary shape, which is similar to our findings. 

Secondly, we find a one-armed SWP that seems to be composed of only the fundamental slow body kink mode ($m = -1$). What is more interesting is that we find two one-armed SWPs coexisting within U1/U2, which can rotate either in the same or opposite directions. We suggest that this phenomenon arises from different driving sources of umbral waves or the excitation of higher-order MHD modes within an umbra. Some studies suggest that multiple sources of waves exist within the sunspot umbra, each of which may have its own subphotospheric oscillation driver \citep{AballeVillero+etal+1993, Liang+etal+2011, Yurchyshyn+etal+2020}. We should treat this conjecture with particular caution, as the presence of multiple driving sources of umbral waves would further complicate sunspot oscillation patterns. 

Finally, by analyzing the spatial distribution of the oscillation centers of the one-armed SWPs within the umbra, we find that the chromospheric umbral waves are repeatedly triggered in regions with high oscillation power, and most of the umbral waves occur in the dark nuclei and strong magnetic field regions of the umbra, rather than in the regions outside the dark nuclei where the magnetoconvection is more intense. Observationally, the dark nuclei exhibit strong magnetic fields where convection is almost completely suppressed \citep{Weiss+2002, Thomas+etal+2004}. According to the observational results of \cite{Braun+etal+1988} and \cite{Braun+etal+1990}, the p-modes absorption coefficient increases with the magnetic field strength, which mutually corroboration our observational characterization. Therefore, our research results are more inclined to support the p-modes driving.

Before this study, several authors had reported evidence of repeated triggering of the umbral flashes at the locations of the lowest umbral intensity \citep{AballeVillero+etal+1993, RouppevanderVoort+etal+2003, Yurchyshyn+etal+2020, Cho+etal+2021}, but the difference from our work is that their object of study was the umbral flashes. Umbral flashes are the periodic brightening phenomenon of about 3 minutes in the chromospheric umbra \citep{Beckers+etal+1969}, caused by the steepening of upward-propagating magnetoacoustic waves into shocks \citep{Tian+etal+2014, Henriques+etal+2015}. The temperature at the shock front is approximately 1000 K higher than the surrounding plasma \citep{delaCruzRodr+etal+2013}. Studies have shown that umbral flashes are followed by a blueshifted umbral wave, and then by a redshifted plasma that returns to the initial state \citep{Bogdan+2000, Bard+etal+2010}. Therefore, the spatial distribution of the origin of umbral waves can also be characterized by the statistics of the umbral flashes. \cite{RouppevanderVoort+etal+2003} reported no correlation between umbral flashes and umbral dots. We note that several discrete umbral flashes may occur during the evolution of a one-armed SWP. Meanwhile, our conclusions are in contradiction with the results of some authors. \cite{Jess+etal+2012} and \cite{Chae+etal+2017} observed the enhancement of the 3-minute oscillation power above the light bridge and the umbral dots. According to the study by \cite{Yurchyshyn+etal+2015}, umbral flashes tend to occur above the light bridge and the umbral dots. \cite{Cho+etal+2019} discovered four umbral oscillation events that originated from umbral dots. \cite{Cho+etal+2021} and \cite{Wu+etal+2023} suggested that the horizontal position of the oscillation center varies with height when slow magnetoacoustic waves propagate along nonvertical magnetic fields. Moreover, we speculate that the absorption of p-modes in the light bridge and the umbral dots may also generate a small amount of umbral waves, which could account for the discrepancy in the observed results.

In this study, we did not find any evidence of superposition of adjacent umbral oscillations. In the chromosphere, the umbra oscillations experienced rapid damping near the umbral boundaries. Upon detailed investigation, we have discerned that the oscillation signals within the umbrae of these two sunspots exhibit a weak correlation (with a maximum correlation coefficient of 0.3), and their dominant oscillation frequencies are also different. Therefore, the analytical process applied to active region NOAA 13023 is equally applicable to other active regions.

The conclusions of this paper can help us to understand the true nature of the driving source of umbral waves and lay the foundation for further characterizing the plasma state and the magnetic field structure. Despite our research has yielded some results, there are still some problems that need further consideration. In our study, the present umbral waves formed a one-armed SWP only about one-third of the time. Therefore, does the spatial distribution of the oscillation centers of the one-armed SWPs represent the spatial distribution of the origin of all umbral waves within the umbra? To fully answer this question, we need to further investigate the spatial distribution of the oscillation centers of the circular ripple-shaped umbral waves. In addition, to reduce the oscillation center position deviation caused by the magnetic field inclination, we suggest selecting sunspots with dark nuclei to investigate the spatial connection between the 3/5-minute oscillation signals and the dark nuclei/umbral dots based on photospheric velocity oscillations. We should also note that probing the number of driving sources of umbral waves provides a clue to distinguish between the monolithic model \citep{Hoyle+1949} and the cluster model \citep{Parker+1979} of sunspots.

%

%

%

%
\begin{acks}
We thank the reviewer for careful reading of the manuscript and for constructive suggestions that improved the original version of the manuscript. We also would like to acknowledge the IRIS, SDO, and GST science teams for providing the data. 
\end{acks}

\appendix   

\section{Extra Material}
\label{material}
Additional Supporting Information may be found in the online version of this article:
\begin{itemize}
\item \texttt{Animation1.mp4}
\item \texttt{Animation2.mp4}
\item \texttt{Animation3.mp4}
\end{itemize}

\begin{table}
	\caption{One-armed SWPs within the umbrae of two sunspots during 17:00-22:45 UT on 29 May 2022.}
	\label{SWPs-table}
	\begin{tabular}{c|ccc|ccc}
		\hline
		& \multicolumn{3}{c|}{Umbra 1}                                                                             & \multicolumn{3}{c}{Umbra 2}                                                                              \\
		\multirow{-2}{*}{One-armed} & \cellcolor[HTML]{FFFFFF}Start time & \cellcolor[HTML]{FFFFFF}End time & \cellcolor[HTML]{FFFFFF}Lifetime & \cellcolor[HTML]{FFFFFF}Start time & \cellcolor[HTML]{FFFFFF}End time & \cellcolor[HTML]{FFFFFF}Lifetime \\
		SWPs                        & (UT)                               & (UT)                             & (s)                            & (UT)                               & (UT)                             & (s)                            \\ \hline
		\rowcolor[HTML]{FFFFFF} 
		1                           & 17:00:51                           & 17:02:53                         & 122                              & 17:01:32                           & 17:02:53                         & 81                               \\
		\rowcolor[HTML]{FFFFFF} 
		2                           & 17:08:17                           & 17:09:38                         & 81                               & 17:03:33                           & 17:05:35                         & 122                              \\
		\rowcolor[HTML]{FFFFFF} 
		3                           & 17:09:58                           & 17:12:00                         & 122                              & 17:05:55                           & 17:08:37                         & 162                              \\
		\rowcolor[HTML]{FFFFFF} 
		4                           & 17:28:52                           & 17:33:15                         & 263                              & 17:18:45                           & 17:20:26                         & 101                              \\
		\rowcolor[HTML]{FFFFFF} 
		5                           & 17:33:36                           & 17:34:57                         & 81                               & 17:43:03                           & 17:44:44                         & 101                              \\
		\rowcolor[HTML]{FFFFFF} 
		6                           & 17:55:12                           & 17:57:13                         & 121                              & 17:49:07                           & 17:50:49                         & 102                              \\
		\rowcolor[HTML]{FFFFFF} 
		7                           & 18:08:02                           & 18:10:44                         & 162                              & 17:49:28                           & 17:50:49                         & 81                               \\
		\rowcolor[HTML]{FFFFFF} 
		8                           & 18:14:06                           & 18:16:08                         & 122                              & 17:58:14                           & 17:59:55                         & 101                              \\
		\rowcolor[HTML]{FFFFFF} 
		9                           & 19:11:09                           & 19:13:31                         & 142                              & 18:06:00                           & 18:08:22                         & 142                              \\
		\rowcolor[HTML]{FFFFFF} 
		10                          & 19:19:35                           & 19:21:37                         & 122                              & 18:17:29                           & 18:20:11                         & 162                              \\
		\rowcolor[HTML]{FFFFFF} 
		11                          & 19:37:08                           & 19:39:50                         & 162                              & 18:20:31                           & 18:22:12                         & 101                              \\
		\rowcolor[HTML]{FFFFFF} 
		12                          & 19:54:41                           & 19:57:23                         & 162                              & 19:04:03                           & 19:06:45                         & 162                              \\
		\rowcolor[HTML]{FFFFFF} 
		13                          & 20:00:05                           & 20:02:07                         & 122                              & 19:11:09                           & 19:13:51                         & 162                              \\
		\rowcolor[HTML]{FFFFFF} 
		14                          & 20:11:34                           & 20:13:15                         & 101                              & 19:41:32                           & 19:43:33                         & 121                              \\
		\rowcolor[HTML]{FFFFFF} 
		15                          & 20:21:42                           & 20:24:03                         & 141                              & 19:49:58                           & 19:51:39                         & 101                              \\
		\rowcolor[HTML]{FFFFFF} 
		16                          & 20:27:46                           & 20:29:48                         & 122                              & 20:13:36                           & 20:14:57                         & 81                               \\
		\rowcolor[HTML]{FFFFFF} 
		17                          & 20:28:27                           & 20:31:49                         & 202                              & 20:15:57                           & 20:17:39                         & 102                              \\
		\rowcolor[HTML]{FFFFFF} 
		18                          & 20:46:20                           & 20:48:22                         & 122                              & 20:20:21                           & 20:22:42                         & 141                              \\
		\rowcolor[HTML]{FFFFFF} 
		19                          & 20:52:04                           & 20:54:26                         & 142                              & 20:42:58                           & 20:44:59                         & 121                              \\
		\rowcolor[HTML]{FFFFFF} 
		20                          & 20:57:49                           & 20:59:10                         & 81                               & 20:54:06                           & 20:55:47                         & 101                              \\
		\rowcolor[HTML]{FFFFFF} 
		21                          & 20:58:49                           & 20:59:50                         & 61                               & 20:56:28                           & 20:58:29                         & 121                              \\
		\rowcolor[HTML]{FFFFFF} 
		22                          & 20:59:50                           & 21:01:31                         & 101                              & 21:22:07                           & 21:23:28                         & 81                               \\
		\rowcolor[HTML]{FFFFFF} 
		23                          & 21:00:10                           & 21:01:52                         & 102                              & 21:28:52                           & 21:31:14                         & 142                              \\
		\rowcolor[HTML]{FFFFFF} 
		24                          & 21:19:05                           & 21:21:47                         & 162                              & 21:50:28                           & 21:52:09                         & 101                              \\
		\rowcolor[HTML]{FFFFFF} 
		25                          & 21:22:47                           & 21:24:49                         & 122                              & 22:07:41                           & 22:09:22                         & 101                              \\
		\rowcolor[HTML]{FFFFFF} 
		26                          & 21:30:13                           & 21:34:36                         & 263                              & 22:21:31                           & 22:23:13                         & 102                              \\
		\rowcolor[HTML]{FFFFFF} 
		27                          & 21:34:56                           & 21:36:38                         & 102                              &                                    &                                  &                                  \\
		\rowcolor[HTML]{FFFFFF} 
		28                          & 21:37:59                           & 21:40:00                         & 121                              &                                    &                                  &                                  \\
		\rowcolor[HTML]{FFFFFF} 
		29                          & 21:54:11                           & 21:56:12                         & 121                              &                                    &                                  &                                  \\
		\rowcolor[HTML]{FFFFFF} 
		30                          & 21:56:33                           & 21:58:54                         & 141                              &                                    &                                  &                                  \\
		\rowcolor[HTML]{FFFFFF} 
		31                          & 22:02:17                           & 22:04:18                         & 121                              &                                    &                                  &                                  \\
		\rowcolor[HTML]{FFFFFF} 
		32                          & 22:07:14                           & 22:10:23                         & 189                              &                                    &                                  &                                  \\
		\rowcolor[HTML]{FFFFFF} 
		33                          & 22:11:03                           & 22:12:45                         & 102                              &                                    &                                  &                                  \\
		\rowcolor[HTML]{FFFFFF} 
		34                          & 22:43:08                           & 22:45:09                         & 121                              &                                    &                                  &                                  \\ \hline
		
	\end{tabular}
\end{table}
\section{Additional Statements}
\begin{authorcontribution}
	Xinsheng Zhang downloaded and analyzed the data and wrote the original manuscript. Jincheng Wang and Zhe Xu provided technical support during the data processing. Xiaoli Yan, Zhike Xue, and Zhe Xu supervised and revised the manuscript. Qiaoling Li, Yang Peng, and Liping Yang participated in the discussion of this research work. Xiaoli Yan directed the whole study and is the corresponding author of this manuscript. All authors have read and agreed to the final version of the manuscript.
\end{authorcontribution}
\begin{fundinginformation}
	This work is supported by the Strategic Priority Research Program of the Chinese Academy of Sciences (XDB0560000), the National Natural Science Foundation of China (12325303, 12003068, 12003064, 11973088, and 11973084), the Yunnan Science Foundation of China (202201AT070194, 202101AT070032, and 202301AT070347), the Key Research and Development Project of Yunnan Province (202003AD150019), the Youth Innovation Promotion Association, Chinese Academy of Sciences (CAS; 2019061), the CAS “Light of West China” Program, the Yunnan Key Laboratory of Solar Physics and Space Science (202205AG070009), the Yunnan Science Foundation for Distinguished Young Scholars (202001AV070004), and the Yunnan Provincial Science and Technology Department (202305AH340002).
\end{fundinginformation}
\begin{dataavailability}
	The IRIS and SDO data used in this study are available through the download interface of SunPy \citep{TheSunPyCommunity+etal+2020}. The data from GST can be requested for download from \url{http://www.bbso.njit.edu/~vayur/NST_catalog/.} 
\end{dataavailability}
\begin{ethics}
	\begin{conflict}
		The authors declare no competing interests.
	\end{conflict}
\end{ethics}

%
%
\bibliographystyle{spr-mp-sola}
\bibliography{REFERENCES} 
 
%
%
%
%

\end{article} 
\end{document}